\documentclass[3p, article]{elsarticle}

\usepackage{framed,multirow}
\usepackage[utf8]{inputenc}

\usepackage{graphicx}

\usepackage{natbib}
\usepackage{textcomp}
\usepackage{amstext}
\usepackage{amssymb}
\usepackage{amsmath}
\usepackage{tikz}
\usepackage{lettrine}
\usepackage[export]{adjustbox}
\usepackage{ragged2e}
\usepackage{subfig}
\usepackage{float}
\usepackage{booktabs}
\usepackage{makecell, multirow}

\newcolumntype{C}[1]{>{\centering\arraybackslash}m{#1}}
\usepackage{url}
\usepackage{lipsum}

\journal{arXiv}

\title{Mean Absolute Directional Loss as a New Loss Function for Machine Learning Problems in Algorithmic Investment Strategies\tnoteref{t1}}
\tnotetext[t1]{This document is the results of the research project funded by IDUB program: BOB-IDUB-622-233/2022 at the University of Warsaw}

\author[1,2]{Jakub Michańków\fnref{fn1}}
\ead{jmichankow@wne.uw.edu.pl}

\author[2]{Paweł Sakowski\fnref{fn2}}
\ead{sakowski@wne.uw.edu.pl}

\author[2]{Robert Ślepaczuk\corref{cor1}%
           \fnref{fn3}}
\ead{rslepaczuk@wne.uw.edu.pl}

\cortext[cor1]{Corresponding author: rslepaczuk@wne.uw.edu.pl}

\fntext[fn1]{ORCID: 0000-0002-0567-6240; \\email: jmichankow@wne.uw.edu.pl}
\fntext[fn2]{ORCID: 0000-0003-3384-3795; \\email: sakowski@wne.uw.edu.pl}
\fntext[fn3]{ORCID: 0000-0001-5227-2014}

\affiliation[1]{
    organization={Department of Informatics,Cracow University of Economics},
    addressline={ul. Rakowicka 27},
    city={Cracow},
    postcode={31-510},
    country={Poland}
}
\affiliation[2]{
    organization={Quantitative Finance Research Group, Department of Quantitative Finance, Faculty of Economic Sciences, University of Warsaw},
    addressline={ul. Długa 44/50},
    postcode={00-241},
    city={Warsaw},
    country={Poland}
    }

\date{September 2023}

\begin{document}

\begin{abstract}
    This paper investigates the issue of an adequate loss function in the optimization of machine learning models used in the forecasting of financial time series for the purpose of algorithmic investment strategies (AIS) construction. We propose the Mean Absolute Directional Loss (MADL) function, solving important problems of classical forecast error functions in extracting information from forecasts to create efficient buy/sell signals in algorithmic investment strategies. Finally, based on the data from two different asset classes (cryptocurrencies: Bitcoin and commodities: Crude Oil), we show that the new loss function enables us to select better hyperparameters for the LSTM model and obtain more efficient investment strategies, with regard to risk-adjusted return metrics on the out-of-sample data.
\end{abstract}


\maketitle

\section{Introduction}

The main idea for this paper comes from the unsolved dilemma regarding the search for forecasting models that can be used for buy/sell signal generation in algorithmic investment strategies (AIS). No matter what kind of theoretical concept is incorporated into the heart of such an investment model, we have a few similar issues that have to be properly addressed to increase the probability of generating efficient signals on out-of-sample (OOS) data. Among many others, these include the architecture of testing various models (machine learning, econometric, macroeconomic, or statistical approaches), the structure of the walk-forward procedure (usually consisting of numerous training, validation, and testing periods of different lengths), hyperparameters tuning and parameters optimization, model estimation phase, and finally the appropriate set of time series with possibly diverse characteristics of their distributions. The point is that all of these problems have to be designed optimally in order to avoid potential over-fitting issues and find the best possible variant of the investment model. 

Majority of papers undertaking the topic of AIS testing do not put proper attention to these problems and focus only on the empirical testing of one or several selected investment models, on a single instrument, over quite short data periods, usually without explaining the details of the whole procedure. In this paper, we decided to focus on one crucial aspect of testing such models, which is, in our opinion, the selection of a proper loss function. In reality, it has the greatest impact on hyper-parameter tuning, followed by the model estimation phase.

The main hypothesis verified in this paper is as follows (RH): \textit{The MADL loss function has better properties than classical forecast error functions in the optimization of ML models used in forecasting financial time series for the purpose of AIS construction}. In order to verify this hypothesis, after a brief reference to the most common drawbacks in papers testing AIS (Literature Review section), and a description of our architecture of testing (Methodology and Data section), we define and describe the new formula of the MADL function, focusing primarily on its conceptual differences to other similar functions. Then, in the Results section, we show the empirical comparison of the use of the MADL function and the classical forecast error function (MAE) on two various time series: Bitcoin and Crude Oil daily simple returns. In this analysis, we visualize potential differences in a more realistic way than only theoretical deliberations. Finally, we conclude and present some extensions of this research.
\section{Literature review}

The literature review presented below focuses on critical drawbacks in papers that describe the testing of algorithmic investment strategies. The main issue is that the vast majority of these papers do not maintain a proper testing structure, which is why their results cannot be treated as valid and robust, despite the fact that the literature on this topic is very broad. Prior to moving on to the main part of this study, it is crucial to list and describe the most frequent flaws in papers examining different AIS. These include:

\begin{itemize}

\item Over-optimization of machine learning models (\cite{lopez2013look}, \cite{bailey2016probability}.

\item Wrong optimization criteria or loss functions, including RMSE, MSE, MAE, MAPE, percent of over-predictions (\%OP), and others. These are used by authors in a vast majority of publications, making it hard to choose the optimal methods for producing buy/sell signals (\cite{dipersioArtificialNeuralNetworks2016}, \cite{yangDeepLearningStock2019}).

\item Only one in-sample period and one out-of-sample period, making the results highly dependent on the chosen period. A great number of research papers on this topic (\cite{wiecki2016all}, \cite{lopez2013look}, \cite{bailey2016probability}, \cite{raudys2016portfolio}) highlight this issue as being very prevalent.

\item Because there was no out-of-sample period, the paper's findings lack any forecasting ability (\cite{TOPCU2020101691}, \cite{CAPORALE2019}.

\item As there is only one basis instrument used for AIS testing, the distributional characteristics of this instrument have a strict influence on the results (\cite{vo2020high}.

\item Forward-looking bias in buy/sell signals, which is typically caused by the use of future macroeconomic data or mistakes in the definitions of buy/sell signals (\cite{chan2013algorithmic}, \cite{chan2021quantitative}, \cite{jansen2020machine})

\item Lack of sensitivity analysis, which is essential in assessing robustness of model's results, with regard to the parameters that were initially chosen (\cite{dipersioArtificialNeuralNetworks2016}, \cite{zhang_multi_2018}, and \cite{yangDeepLearningStock2019}.

\item Data snooping bias - when authors publish only their best results, without conducting a systematic examination of other parameters and assumptions(\cite{bailey2016backtest}, \cite{chan2013algorithmic}).

\item Survivorship bias - the selection of current index constituents, for instance, in research using data from the previous 20 years, is the most prevalent illustration of this bias (\cite{chan2021quantitative}).

\item Inadequate performance metrics or their incorrect interpretation - the efficiency of investment strategies is evaluated only on the quality of point forecasts generated using the theoretical models that form the basis of such investment strategies, rather than of proper risk-adjusted return metrics calculated on the equity lines generated by these strategies (\cite{CHAKOLE2021113761}, \cite{GROBYS2020101396}).

\end{itemize}

Some of the above mentioned drawbacks can be mitigated, at least in part, by pertinent testing of the model architecture. Special attention has to be paid to the hyperparameter tuning phase, where an appropriately selected loss function is critical. 

\section{Methodology and Data}

\subsection{MADL. New Loss Function}\label{MADL. New Loss Function} 

We introduce our novel loss function to address one of the most prevalent issues with papers testing algorithmic trading strategies. In this study, we use it to enhance the predictive power of the LSTM model (\cite{hochreiter_long_1997}) for these strategies.

Based on prior studies (e.g. \cite{voslepaczuk2021}), we came to the conclusion that popular error metrics like RMSE, MSE, MAE, MAPE, and \%OP used in the majority of similar studies are not appropriate error functions for assessing the effectiveness of the models' forecasting abilities in AIS. The reason for this is that the error metrics mentioned above do not consider the forecasting ability of investment signals that are based on these forecasts; rather, they only consider the forecasting accuracy of forecasts (i.e., the difference between the forecasted and observed value). It implies that almost all of these error metrics (RMSE, MSE, MAE, and MAPE) penalize forecast errors regardless of whether they are positive or negative (\(\textrm{forecast error}=\hat{R}_{i}-R_{i}\)), whereas the \%OP metric only considers the forecast error's direction and not its magnitude. For this reason, researchers in the majority of other papers choose the signal combination that optimizes only the chosen error metric, rather than the most profitable combination of signals for the strategy.

To solve this problem, we suggest a new loss function, called Mean Absolute Directional Loss (MADL), which can be computed using the formula below:

\begin{equation}
\textrm{MADL} =\frac{1}{N} \sum_{i=1}^{N} (-1) \times \textrm{sign}({R_{i} \times \hat{R}_{i}}) \times  \textrm{abs}(R_{i} ),
\end{equation}
where \textrm{MADL} is the Mean Absolute Directional
Loss, $R_{i}$ is the observed return on interval \(i\), \(\hat{R}_{i}\) is the predicted return on the interval $i$, $\textrm{sign}(X)$ is the function that returns -1,0,1 as the sign of $X$, $\textrm{abs}(X)$ is the function that gives the absolute value of $X$ and $N$ is the number of forecasts. In this manner, the value the function returns will be equal to the observed return on investment with the predicted direction, allowing the model to determine whether the prediction will result in profit or loss and the amount of this profit or loss. MADL was designed specifically for generating signals for AIS. In our model, this function is minimized, ensuring that the strategy will produce a profit if it returns negative values and loss if it returns positive values. MADL was also the main loss function used in hyperparameters tuning and in the estimation of the LSTM~model.

\begin{figure}[H]
\centering
\includegraphics[width=1\textwidth]{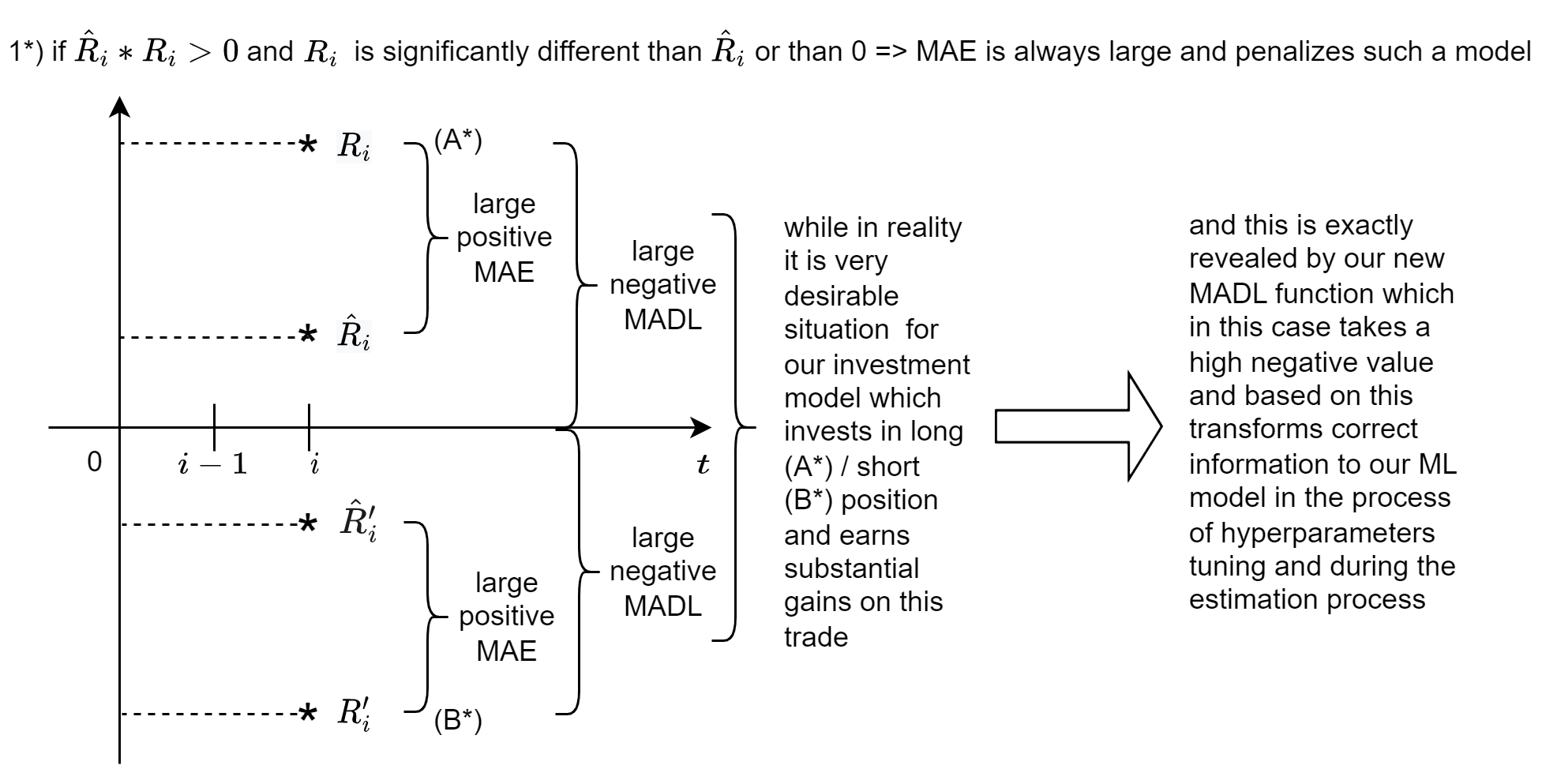}
\footnotesize
\justifying
Note: Variants (A*), and (B*) show that in specific cases the difference between MAE and MADL values can be significant, and therefore affect the final selection of model parameters.
\caption{The difference between MAE and MADL logic \#1.}
\label{fig:MADLconcept-1}
\end{figure}

\begin{figure}[H]
\centering
\includegraphics[width=1\textwidth]{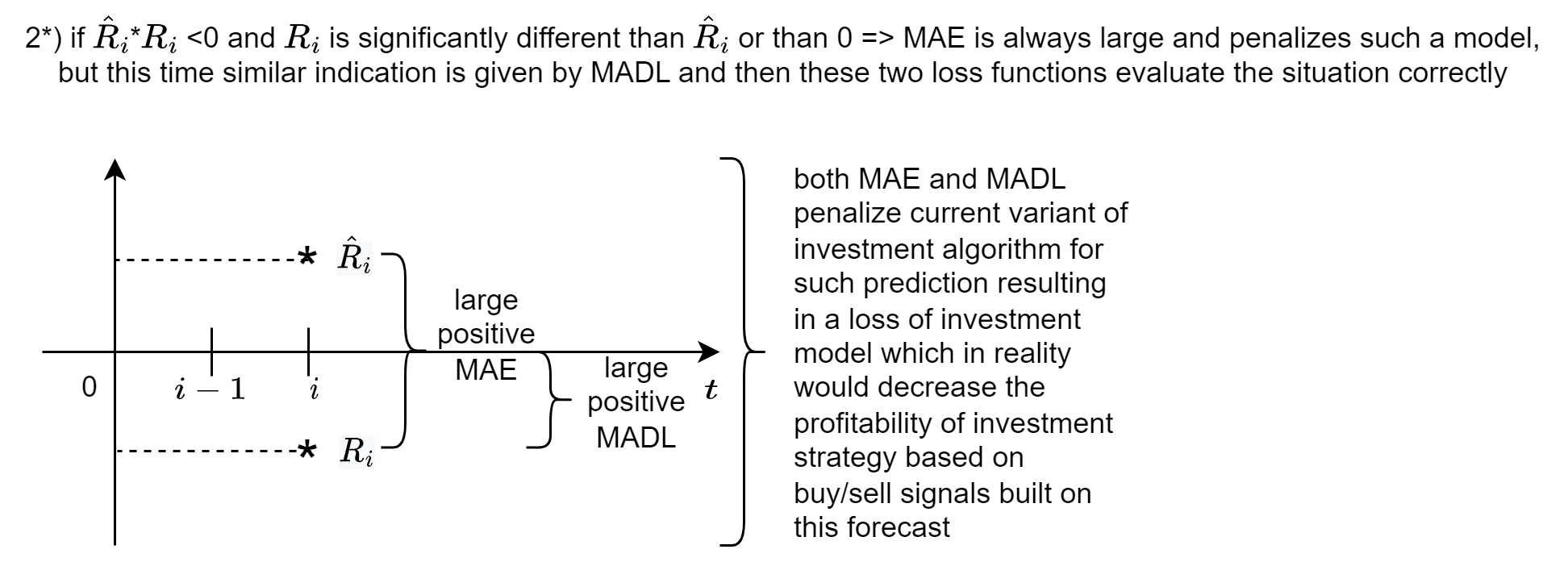}
\footnotesize
\justifying
Note: This variant depicts the difference between MAE and MADL by treating the error as the distance between the predicted and observed values (MAE) and the observed and 0 (MADL) in cases where the sign of the prediction is not the same.
\caption{The difference between MAE and MADL logic \#2.}
\label{fig:MADLconcept-2}
\end{figure}

\begin{figure}[H]
\centering
\includegraphics[width=1\textwidth]{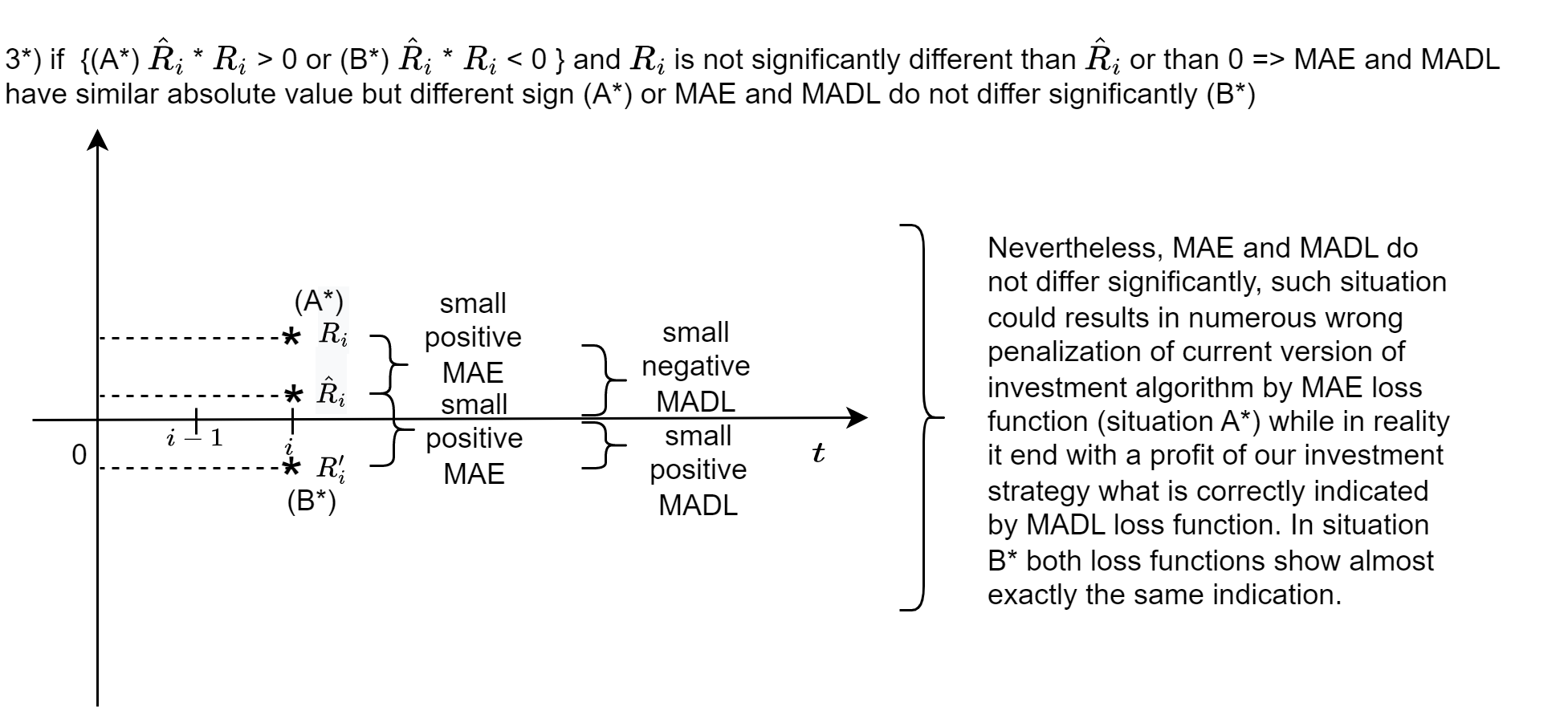}
\footnotesize
\justifying
Note: The last example of the difference is negligible with regard to its value for a single prediction, but taking into account that we can have quite a few of them, it can substantially influence the aggregated value of the selected loss function. 
\caption{The difference between MAE and MADL logic \#3.}
\label{fig:MADLconcept-3}
\end{figure}

In order to reveal the properties of the MADL function and visualize its differences with regard to classical error metrics (e.g. MAE) we present its main distinguishing characteristics, shown in three consecutive figures (Figure~\ref{fig:MADLconcept-1}, ~\ref{fig:MADLconcept-2}, and ~\ref{fig:MADLconcept-3}).

Figure \ref{fig:MADLconcept-1} depicts the difference between MAE and MADL logic that is responsible for the large differences between these two loss functions.

Figure \ref{fig:MADLconcept-2} presents the difference between MAE and MADL logic in the case when the difference is relatively small but still affects the final values of the loss functions.

Figure \ref{fig:MADLconcept-3} presents the difference between MAE and MADL logic in the case when predicted and observed returns have an opposite sign and their values are significantly different.

\begin{figure}[H]
\includegraphics[width=1\textwidth]{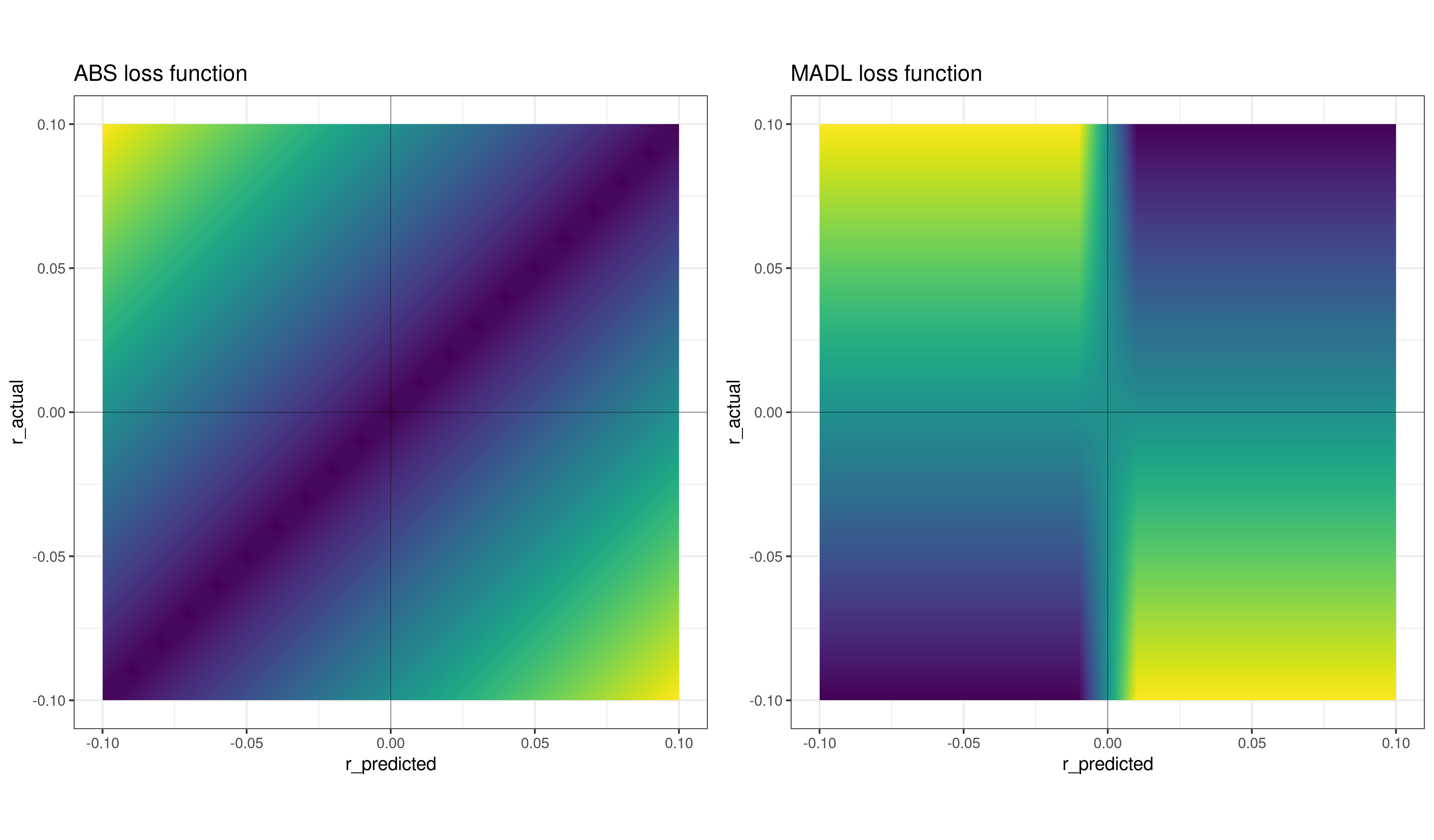}
\footnotesize 
\justifying
Note: Dark blue color indicates the lowest error, while yellow color indicates the highest error. The left panel (with MAE) shows that the error is strictly dependent on the difference between \(\hat{R}_{i}\) and \(R_{i}\) while the right panel shows that \(\hat{R}_{i}\) is always compared with the zero reference point in the process of error calculation.
\caption{Loss functions: MADL vs MAE}
\label{fig:madl_vs_mae}
\end{figure}

Figures (Figure~\ref{fig:MADLconcept-1}, ~\ref{fig:MADLconcept-2}, and ~\ref{fig:MADLconcept-3}) show the most important conceptual differences between classical error functions and MADL. Additionally, Figure~\ref{fig:madl_vs_mae} presents the distribution of error in cases of large and small differences of observed return (\(R_{i}\)) from predicted return (\(\hat{R}_{i}\)) or from zero. What is significant about the MADL loss function is that the reference point for loss measurement is set to zero, whereas, in almost all other classical error metrics, it is the predicted return (\(\hat{R}_{i}\)). Figure \ref{fig:madl_vs_mae} presents the comparison of MAE and MADL for various values of predicted and observed returns.

\subsection{Architecture of testing and research description}\label{Terminology and Metrics and research description}

\begin{figure}[H]
\centering
\includegraphics[width=0.99\textwidth]{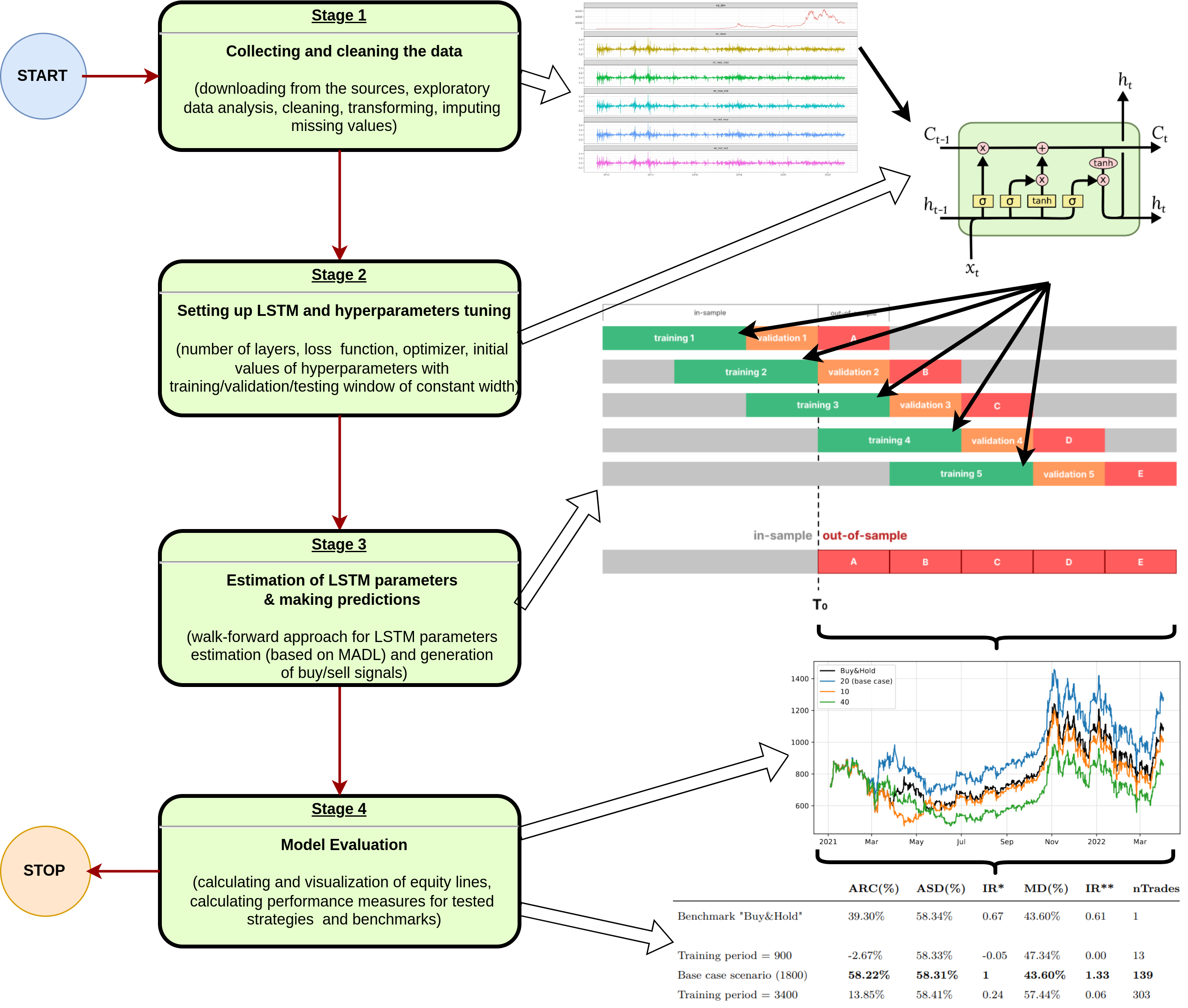}
\footnotesize
Note: MADL function is used in stages 2 and 3.
\caption{The process of LSTM model testing with MADL function}
\label{fig:A-of-T}
\end{figure}

Before we move to the empirical part, it is important to describe the proper architecture for testing ML models used for AIS construction. Although the main aim of this paper is the presentation of the concept and main characteristics of the new loss function, the empirical part (see Section \ref{Results}) with the testing example is added for visualization purposes and as an explanation of its use in practice. Figure \ref{fig:A-of-T} presents the process used in order to explain how exactly our AIS was tested.

We used the LSTM model with a walk-forward procedure of testing. The hyperparameters tuning phase was performed on the first in-sample period, then we trained and estimated our model on the remaining data. For this purpose, we used various combinations of the loss functions in order to extract information about the usefulness of the selected loss function in such a procedure. Table \ref{tab:params} summarizes the values of hyperparameters selected in the tuning phase for each loss function separately.

\begin{table}[H]
\caption{Selected values of hyperparameters.}
\label{tab:params}
\centering\fontsize{8}{8}\selectfont
\resizebox{16.5cm}{!}{
\begin{tabular}{c c c}
\toprule
\textbf{Hyperparameter} & \textbf{Selected Value MADL} & \textbf{Selected Value MAE} \\
\midrule
\hspace{1em} No. hidden layers & 3 & 3  \\
\hspace{1em} No. neurons & 512/256/128 & 64/32/16 \\
\hspace{1em} Activation function & tanh  & tanh  \\
\hspace{1em} Recurrent Activation & sigmoid & sigmoid   \\
\hspace{1em} Dropout rate & 0.02 & 0.0002 \\
\hspace{1em} l2 regularizer & 0.0005 & 0.00001 \\
\hspace{1em} Optimizer & Adam & Adam \\
\hspace{1em} Learning rate & 2.15 & 0.0015 \\
\hspace{1em} BTC train/test & 1460/365 & 1460/365 \\
\hspace{1em} USO train/test & 1008/252 & 1008/252   \\
\hspace{1em} Batch size (BTC/USO) & 1460/1008 & 1460/1008 \\
\hspace{1em} Sequence length (BTC/USO) & 20/10 & 20/10 \\
\hspace{1em} Epochs & 300 & 200 \\
\bottomrule
\end{tabular}
}
\footnotesize \justifying

Note: MADL hyperparameters (second column) were used for MADL/MADL and MADL/MAE strategies, while MAE hyperparameters were used for MAE/MAE and MAE/MADL strategies.
\end{table}

Finally, we performed the following steps in order to perform the empirical part of this research:
\begin{itemize}
\item
  The division of data into in-sample (training and validation) and
  out-of-sample (test) sets, each with 1460/365 observations for BTC and 1008/252 observations for Crude Oil.
\item
  Hyperparameters tuning based on new Loss function: MADL, described in Section \ref{MADL. New Loss Function}, and the old concept of MAE, described above.
\item
  Buy/Sell signals were generated based on the sign of next-day forecasts.
\item
  Tests for one type of strategy: Long/Short.
\item
  Walk-forward approach was used for testing.
\item
  Equity lines and performance metrics from the two loss functions and assets according to \cite{slepaczuk_investment_2018}, results provided in Section \ref{Results}.
\end{itemize}

\subsection{Performance~Metrics}\label{performance-metrics}

In order to evaluate the efficiency of tested strategies, we calculate the following performance metrics based on \cite{kosc_momentum_2019} and \cite{BUI2021126784}. The performance metrics were divided into four categories:

\begin{itemize}

\item return performance metrics: Annualized Return Compounded (ARC)

\item risk performance metrics: Annualized Standard Deviation (ASD), Maximum Drawdown (MD), Maximum Loss Duration (MLD)

\item risk-adjusted return performance metrics: Information Ratio (IR*), Modified Information Ratio (IR**), Aggregated Information Ratio (IR***)

\item informative performance metrics: Number of observations (nObs), Number of trades (nTrades).

\end{itemize}

\subsection{Data}\label{Data}

We decided to use simple returns based on daily data of two assets from various asset classes: Bitcoin (BTC) and Crude Oil (represented by ETF named USO). The data we used covered the period from April 1, 2023 to December 31, 2021. Source of the BTC data: Kraken, Bitfinex, BTC-e, CEX, and Coinbase exchanges, while USO data was obtained from stooq.com.

\section{Results}
\label{Results}

Table \ref{tab:results_cnn_mse2} presents the results of the LSTM model used in this research, for which the hyperparameters tuning, training, and estimation were performed based on various combinations of loss functions: MADL/MADL, MADL/MAE, MAE/MAE, and MAE/MADL. The performance metrics for final equity lines show that no matter which risk-adjusted return metrics (IR*, IR**, and IR***) we select, the result is always the best in s of MADL/MADL approach.

\begin{table}[H]
\centering\fontsize{8}{8}\selectfont
\caption{Comparison of two approaches}
\label{tab:results_cnn_mse2}

\resizebox{16.5cm}{!}{
\begin{tabular}{lccccccccc}
    \toprule
    Model & aRC & aSD & MD & MLD & IR* & IR** & IR*** & nObs & nTrades\\
    \midrule
  \multicolumn{10}{c}{Panel A - BTC} \\
  \midrule
B\&H & 91.25 & 87.37 & 86.67 & 3.24 & 1.04 & 1.100 & 0.310 & 4107 & 2\\
MADL/MADL & 109.94 & 87.34 & 75.04 & 1.91 & 1.26 & 1.844 & 1.062 & 4107 & 306\\
MADL/MAE & 99.90 & 87.36 & 86.67 & 3.24 & 1.14 & 1.318 & 0.406 & 4107 & 4\\
MAE/MAE & 44.26 & 87.45 & 94.18 & 3.46 & 0.51 & 0.238 & 0.030 & 4107 & 330\\
MAE/MADL & 4.05 & 87.51 & 90.99 & 4.83 & 0.05 & 0.002 & 0.000 & 4107 & 320\\
  \midrule
  \multicolumn{10}{c}{Panel B - USO} \\
  \midrule
  BH & -10.59 & 38.09 & 98.19 & 14.24 & -0.28 & -0.030 & 0.000 & 3901 & 2\\
MADL/MADL & 5.25 & 38.08 & 62.73 & 7.58 & 0.14 & 0.012 & 0.000 & 3901 & 731\\
MADL/MAE & -22.70 & 38.07 & 98.91 & 15.42 & -0.60 & -0.137 & -0.002 & 3901 & 11\\
MAE/MAE & -17.02 & 38.08 & 96.88 & 15.42 & -0.45 & -0.079 & -0.001 & 3901 & 402\\
MAE/MADL & 4.19 & 38.08 & 84.39 & 12.74 & 0.11 & 0.005 & 0.000 & 3901 & 692\\

  \bottomrule
\end{tabular}
}
\footnotesize \justifying

Note: Panel A presents results for BTC, while panel B present the results for USO.
\end{table}

\begin{figure}[H]
\includegraphics[width=1\textwidth]{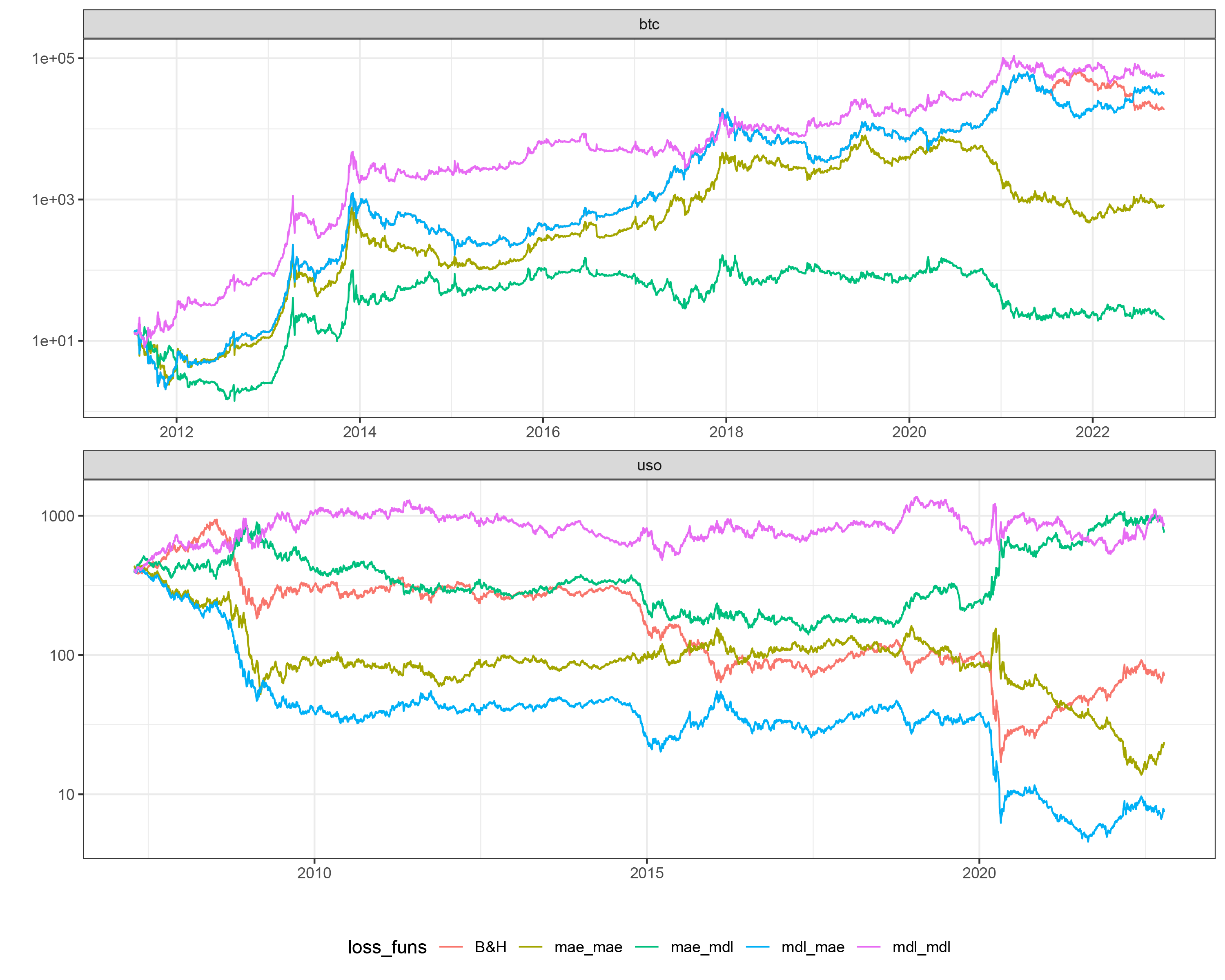}
\footnotesize
Note: Equity lines present the fluctuations of investment strategies for BTC (upper panel) and USO (lower panel) for strategies trained based on various combinations of loss functions used (MAE/MAE, MAE/MADL, MADL/MAE, MADL/MADL) in the period between April 1, 2023, and December 31, 2021. Additionally, we included Buy \& Hold (B\&H) strategy as a benchmark strategy for comparison purposes.
\caption{Equity lines.}
\label{fig:equity_lines}
\end{figure}

Figure \ref{fig:equity_lines} presents the fluctuations of equity lines of for BTC and USO in two panels (top: BTC and bottom: USO). Figure \ref{fig:equity_lines} which shows the best results for MADL/MADL can be treated as an additional confirmation of numeric results presented in Table \ref{tab:results_cnn_mse2}.

Table \ref{tab:regresssion_results} presents the results of a test of significance of $\alpha$ and $\beta$ coefficients from the regression in the form of $R_t = \alpha + \beta R^*_t + \varepsilon_t$, where $R_t$ is the buy and hold returns, and~$R^*_t$ returns from Long/Short strategy. The~results presented in Table~\ref{tab:regresssion_results} confirm those presented in Table~\ref{tab:results_cnn_mse2}, with positive and significant $\alpha$ for MADL/MADL strategy in case of BTC and confirm only partly MADL/MADL strategy for USO (positive $\alpha$ but not significant).

\begin{table}[!h]
\centering\fontsize{8}{8}\selectfont
\caption{Results of regressions for returns: B\&H vs. MAE/MAE, MAE/MADL, MADL/MAE, MADL/MADL.}
\label{tab:regresssion_results}
\resizebox{16.5cm}{!}{
\begin{tabular}[t]{lrrrrrrrr}
\toprule
Dep. var. & Alpha & StdErr & $t$ & pv & Beta & StdErr & $t$ & pv\\
\midrule
\multicolumn{9}{c}{Panel A - BTC}\\
\midrule
\hspace{0em}MAE/MAE & 0.0009 & 0.001 & 1.431 & 0.152 & 0.394 & 0.014 & 27.4 & 0.000\\
\hspace{0em}MAE/MADL & 0.0007 & 0.001 & 1.014 & 0.310 & 0.159 & 0.015 & 10.2 & 0.000\\
\hspace{0em}MADL/MAE & 0.0004 & 0.000 & 1.307 & 0.191 & 0.895 & 0.007 & 129.0 & 0.000\\
\hspace{0em}MADL/MADL & 0.0029 & 0.001 & 4.027 & 0.000 & 0.073 & 0.016 & 4.6 & 0.000\\
\midrule
\multicolumn{9}{c}{Panel B - USO}\\
\midrule
\hspace{0em}MAE/MAE & -0.0005 & 0.000 & -1.423 & 0.154 & -0.374 & 0.015 & -25.1 & 0.000\\
\hspace{0em}MAE/MADL & 0.0004  & 0.000 & 1.104  & 0.269 & -0.236 & 0.016 & -15.1 & 0.000\\
\hspace{0em}MADL/MAE & -0.0006 & 0.000 & -1.956 & 0.050 &  0.491 & 0.014 &  35.1 & 0.000\\
\hspace{0em}MADL/MADL & 0.0004  & 0.000 & 1.215  & 0.224 & -0.228 & 0.016 & -14.6 & 0.000\\
\bottomrule
\end{tabular}
}
\justifying 
{\footnotesize 

Note: The~table presents the results of regressions in the form of: $R_t = \alpha + \beta R^*_t + \varepsilon_t$, where $R_t$ is the return for tested strategy in period $t$ and $R^*_t$ is the return in of BTC or USO strategies. Regressions were calculated in the period between April 1, 2023 to December 31, 2021. The~hyperparameters of LSTM model for the base case scenario were set as it was described in Table~\ref{tab:params}.}
\end{table}

\section{Conclusions}

This paper aimed to introduce the new concept of a more accurate loss function, which is appropriately adjusted to the problem of the optimization of machine learning models used in the forecasting of financial time series, and for construction of algorithmic investment strategies (AIS). For this purpose, we proposed a new loss function -- MADL (Mean Absolute Directional Loss), presented the rationale for its construction, and a detailed interpretation with special attention to its difference with regard to classical loss functions used in the literature. Last but not least, we demonstrated how the new loss function helps us choose better hyperparameters for our ML models and produce more effective investment strategies in terms of risk-adjusted return metrics on the out-of-sample data using two different asset classes (cryptocurrencies: Bitcoin and commodities: Crude Oil).

In the next step we intend to modify the original MADL loss function introducing various penalization structures, e.g. "squared" instead of "absolute" which could stress the problem of large losses in the process of optimization. Such modification could help us with training of various ML models responsible for forecasts used in the buy/sell signals generation process.

\bibliography{bibfile1, bibfile2, bibfile3}{}

\end{document}